\documentclass{article}

\usepackage{PRIMEarxiv}

\usepackage[utf8]{inputenc} % allow utf-8 input
\usepackage[T1]{fontenc}    % use 8-bit T1 fonts
\usepackage{hyperref}       % hyperlinks
\usepackage{url}            % simple URL typesetting
\usepackage{booktabs}       % professional-quality tables
\usepackage{amsfonts}       % blackboard math symbols
\usepackage{nicefrac}       % compact symbols for 1/2, etc.
\usepackage{microtype}      % microtypography
\usepackage{times}
\usepackage{fancyhdr}       % header
\usepackage{graphicx}       % graphics
\usepackage{indentfirst}
\usepackage{pgfplots}
\usepackage{tabularx}
\usepackage{booktabs}
\usepackage{pdflscape}
\usepackage{longtable}
\usepackage{tikz}
\usepackage{scrextend}
\graphicspath{{media/}}     % organize your images and other figures under media/ folder

\title{A Survey of Large Language Models in Cybersecurity}

\author{
  Gabriel de Jesus Coelho da Silva \\
  Master's Researcher - Computer Science \\
  Universidade Federal de Santa Catarina - UFSC \\
  Florianópolis, SC, Brazil \\
  \texttt{gabriel.jcs@posgrad.ufsc.br} \\
   \And
  Carlos Becker Westphall \\
  Full Professor - Network and Management Laboratory \\
  Universidade Federal de Santa Catarina - UFSC \\
  Florianópolis, SC, Brazil \\
  \texttt{carlosbwestphall@gmail.com} \\
}

\begin{document}
\maketitle

\begin{abstract}
Large Language Models (LLMs) have quickly risen to prominence due to their ability to perform at or close to the state-of-the-art in a variety of fields while handling natural language. An important field of research is the application of such models at the cybersecurity context. This survey aims to identify where in the field of cybersecurity LLMs have already been applied, the ways in which they are being used and their limitations in the field. Finally, suggestions are made on how to improve such limitations and what can be expected from these systems once these limitations are overcome.
\end{abstract}

\section{Introduction}

\subsection{Motivation}

The cybersecurity industry is a field that, by its nature, requires keeping up with the state-of-the-art constantly, given that every new technology provides new avenues for malicious exploitation and, as such, may increase the available attack surface. 

Software vulnerabilities are errors introduced in software, accidentally or deliberately, allowing threat actors to exploit the behavior of a system in a harmful way that is not intended or expected by its developers. Software vulnerabilities exploitation may lead to user data exposure or corruption, denial-of-service, or code execution that can lead to the complete takeover of a system \cite{Bojanova2023-hc}. As such, companies commonly conduct their own controlled attempts at invading their own systems in an effort to detect and fix any vulnerabilities before malicious actors \cite{Teichmann2023}, commonly referred to as a \textbf{vulnerability assessment}, or, interchangeably, red teaming and/or penetration testing.

The evolution of artificial intelligence and its effective usage in a wide variety of fields as a generic tool has also made its way to cybersecurity, both in the offensive as well as defensive side, and its application has become the norm in most commercial and academic settings \cite{Li2018, Zhang2021}.

\subsection{Justifications}

The rise of neural networks and deep learning have led to the development of AI-based malware and intrusion detection and prevention by analyzing deviance from expected patterns or binary features which may indicate malware \cite{Harer2018-ay, Lin2020-up, Liu2012-hv, LOMIO2022111283, a14080224, Wang2020-fd, Sarker2021-gz, Kuzlu2021-bi}. Although these approaches have been successful and are deployed commercially, they still present issues of too many false positives and lack of explainability and are not well suited for more complex tasks \cite{Taddeo2019-av}.

Recent advances in neural network architectures and the introduction of Large Language Models (LLMs) have shown a great capability to generalize beyond their original pre-trained settings \cite{multitask}. Foundation LLMs provide a good starting point as they're pre-trained on a broad language corpus which includes conversational, scientific, question and answer, and code databases where further fine-tuning could be beneficial for the model on such tasks \cite{llama, gpt4}.

The high specificity and advanced skill set necessaries to perform a vulnerability assessment make it a difficult, albeit desired, task to automate \cite{9229752, MCKINNEL2019175, DBLP:journals/corr/abs-1905-05965}.

\subsection{Objectives}

\subsubsection{General objectives}
To survey the usage of LLMs in the context of cybersecurity, understanding if there are any areas in which they excel or need to improve on, and the methods by which their use is achieved.

\subsubsection{Specific objectives}
\begin{itemize}
    \item Understand how LLMs are used in cybersecurity;
    \item Identify how vulnerability assessments are performed with LLMs;
    \item Suggest ways in which improvement could be achieved with LLMs being used as vulnerability agents.
\end{itemize}

\subsection{Paper organization}

An overview of the basic concepts is first provided, outlining both the fields of cybersecurity and artificial intelligence and their intersections.

Next, a survey is carried by finding papers where LLMs are used for cybersecurity applications, the ways in which they're used and the areas in which they are well suited or need to improve upon.

Finally, the paper suggests a new approach through which the usage of LLMs could contribute to the cybersecurity field in an attempt to address the limitations of current systems and where this new approach can lead to.

\section{Basic concepts}

\subsection{Cybersecurity}

The cybersecurity field deals with protecting from and responding to threats in computer systems and its related networks and components. Modern infrastructure is increasingly reliant on computer systems and both industrial as well as home devices are increasingly connected to the internet, meaning protecting systems from outside tempering has become a must
\cite{Kemmerer2003-lo}.

Cyber threats, a diverse array of malicious activities, pose significant risks to individuals, organizations, and even nations. Common threats include malware, which encompasses viruses, worms, and ransomware, as well as phishing attacks that exploit human vulnerabilities through deceptive means. Understanding the various attack vectors, such as exploiting software vulnerabilities, social engineering tactics, and physical breaches, is essential in crafting effective cybersecurity strategies \cite{Humayun2020}.

To counteract these threats, cybersecurity relies on a combination of preventive, detective, and corrective measures. Encryption, firewalls, antivirus software, and other tools serve as key components of a comprehensive cybersecurity toolkit. In addition to technological safeguards, the establishment and enforcement of security policies and procedures, as well as rigorous risk management practices, are crucial in creating a resilient security posture \cite{7371499}. Organizations also need to be cognizant of regulatory frameworks, as various industries and regions may have specific cybersecurity requirements.

As technology evolves, so do the challenges and opportunities in cybersecurity. Emerging trends, such as the integration of artificial intelligence and machine learning \cite{Harer2018-ay, Lin2020-up, Liu2012-hv, LOMIO2022111283, a14080224, Wang2020-fd, Sarker2021-gz, Kuzlu2021-bi}, are reshaping the landscape, offering both new solutions and potential vulnerabilities. Ethical considerations, including privacy concerns and responsible disclosure of vulnerabilities, add an additional layer of complexity to the field.

In the proactive pursuit of cybersecurity excellence, organizations often employ penetration testing and vulnerability assessments as indispensable tools to fortify their defenses. Penetration testing, commonly known as ethical hacking, involves simulated cyber-attacks conducted by skilled professionals to identify and exploit potential vulnerabilities in systems, networks, and applications. This hands-on approach allows organizations to assess the resilience of their security measures in a controlled environment, uncovering weaknesses that malicious actors might exploit \cite{bacudio2011overview, shah2015overview}. Vulnerability assessments, on the other hand, focus on systematically identifying, quantifying, and prioritizing vulnerabilities within an IT environment. By conducting regular assessments, organizations can stay ahead of emerging threats, address weaknesses promptly, and enhance their overall security posture \cite{shinde2016cyber}. These proactive measures play a vital role in maintaining a dynamic and adaptive defense against the evolving landscape of cyber threats.

\subsection{Neural Networks}

Artificial neural networks were first introduced as a mathematical model of the biological neurons in an attempt to develop a machine with the same capabilities of a biological brain \cite{Rosenblatt_1957_6098, Sanger1958ThePA}. The basic neuron is composed of a set of inputs with weights, outputs and an activation function; a network is constructed by the interconnection of neurons. Although with modest results at first, further advances on learning algorithms \cite{rumelhartLearningRepresentationsBackpropagating1986, plautLearningSetsFilters1987} and the stacking of several layers \cite{hintonLearningMultipleLayers2007}, combined with larger memories, more data availability and faster computers have propelled neural networks as a somewhat generic algorithm, capable of reaching the state-of-the-art in a wide variety of fields and applications, as long as there is enough data available for training. This approach is often called "Deep Learning" and has become ubiquitous in the field \cite{lecunDeepLearning2015}.

The most common approach with neural networks is through supervised learning. In this approach, the network goes through an initial training, where previously annotated data is shown with its desired output and the weights are adjusted for the entire dataset. This produces a set of weights which, when deployed, will allow correct labeling of previously unseen data. The performance of the network is related to the quality of the data used to train as well as the amount of training it has gone through.

\subsection{Natural Language Processing and Large Language Models}

Natural Language Processing (NLP) is a field that deals with the understanding of natural language by computing machines \cite{https://doi.org/10.48550/arxiv.1510.00726}. Although other approaches are also possible, the neural network approach has also reached state-of-the-art in NLP tasks and has become the preferred method.

The introduction of the transformer architecture and its attention mechanism \cite{vaswaniAttentionAllYou2023} greatly advanced the field and led to the creation of large language models (LLMs) \cite{llama, gpt4}. LLMs take a range of parameters and are trained in extremely large and varied datasets. These datasets may include questions and answers from websites, encyclopedias, books, source codes and so on. Training is done through "next-word prediction", in which the network is shown a phrase with a missing word or sequence of words and has to correctly identify the missing piece.

In general, the training process involves the following steps:

\begin{labeling}{\textbf{Supervised fine-tuning}}
    \item[\textbf{Tokenization}]
    The data is pre-processed into numerical tokens representing the smallest possible textual data point carrying meaning. 
    This step is also required for inference.
    \item[\textbf{Unsupervised learning}]
    The models are trained to predict the next token from a sequence of tokens in the training data. This creates a "base" model from which further training may or may not be performed.
    \item[\textbf{Supervised fine-tuning}]
    The model receives further training by receiving demonstrations of desired request/response pairs created by humans.
\end{labeling}

 Because the training process requires such a huge amount of memory and processing power, it remains available only to a few big industry players which can manage the computing power necessary to perform this task. However, the increased competition and possibility of contribution from the community has lead to the creation of openly available language models, generally referred to as foundation models. These models have already gone through the training process and are provided as a set of weights, generally with varying parameter sizes so they can be run in more accessible consumer hardware. Further specialization of such models is possible by fine-tuning \cite{dettmersQLoRAEfficientFinetuning2023, huLORALOWRANKADAPTATION, zhangLLaMAAdapterEfficientFinetuning} or specific prompting \cite{radfordLanguageModelsAre, https://doi.org/10.48550/arxiv.2206.07682, weiChainofThoughtPromptingElicits, brownLanguageModelsAre2020}.

\section{Related work}

Deep neural networks are already used extensively in cybersecurity for threat and vulnerability detection \cite{Harer2018-ay, Lin2020-up, Liu2012-hv, LOMIO2022111283}, network intrusion prevention \cite{a14080224, Wang2020-fd}, password guessing \cite{passgan}, and others \cite{Sarker2021-gz, Kuzlu2021-bi}, albeit lack of explainability and excess of false positives undermine trust in these systems \cite{Taddeo2019-av}.

The introduction of the transformer architecture \cite{transformer} and subsequent construction of large language models trained on a varied corpus of text data have shown a remarkable ability for further model specialization \cite{multitask}. Fine-tuning of pre-trained foundation models on task-specific data achieves near or better than state-of-the-art performance when compared to other techniques of machine learning \cite{llama, gpt4, instructgpt}. Fine-tuning large language models on publicly available code databases has led to better performance in code writing \cite{codex, codebert, gen-code}, vulnerable code fixing \cite{automatic-repair, vulrepair}, and finding vulnerable code \cite{transformer-vuln}.

Large language models have also been trained to navigate a browser to look for answers \cite{webgpt} and use external tools through Application Programming Interfaces (APIs) \cite{toolformer, qin2023tool}. Autonomous agents with a mix of tool usage and code execution capabilities have also been shown to perform well \cite{emer-scien}.

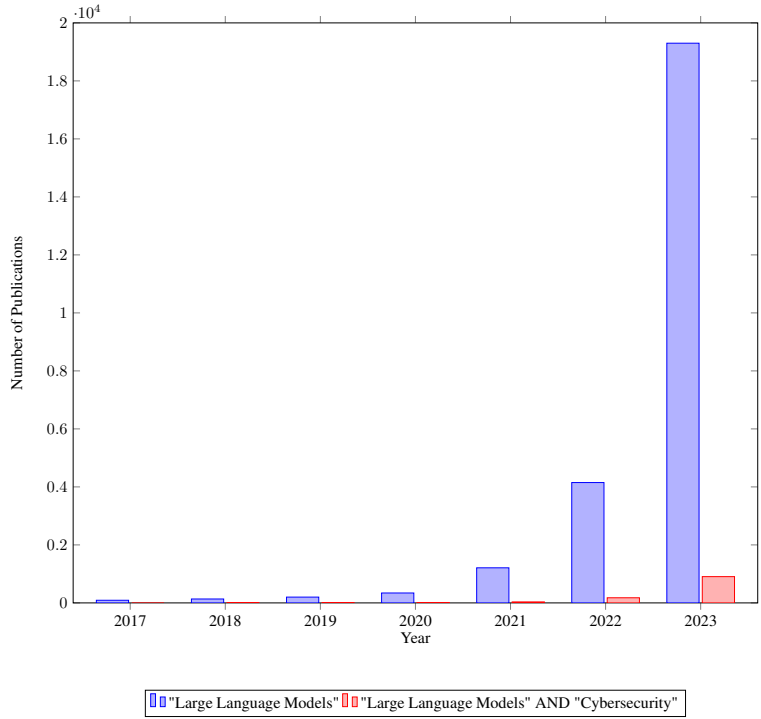
\begin{figure}[h!]
    \centering
\resizebox{10cm}{!}{\begin{tikzpicture}
\begin{axis}[
    ybar,
    width=\textwidth,
    xlabel={Year},
    ylabel={Number of Publications},
    ymin=0,
    ymax=20000, % Adjust this according to your data
    xtick=data,
    xticklabels={2016, 2017, 2018, 2019, 2020, 2021, 2022, 2023},
    bar width=20pt,
    legend style={at={(0.5,-0.15)}, anchor=north, legend columns=-1},
    symbolic x coords={2017, 2018, 2019, 2020, 2021, 2022, 2023},
    every node near coord/.append style={rotate=90, anchor=west},
    nodes near coords align={vertical},
]

\addplot coordinates {(2017, 91) (2018, 135) (2019, 200) (2020, 342) (2021, 1210) (2022, 4150) (2023, 19300)};
\addplot coordinates {(2017, 0) (2018, 1) (2019, 7) (2020, 14) (2021, 34) (2022, 177) (2023, 906)};

\legend{"Large Language Models", "Large Language Models" AND "Cybersecurity"}

\end{axis}
\end{tikzpicture}}
    \caption{Number of publications per year per keyword}
    \label{fig:graphcompare}
\end{figure}

In order to analyze interest in and relevance of LLMs over time, a search of the keywords \textit{"Large Language Models"} and \textit{"Large Language Models" AND "Cybersecurity"} was performed on Google Scholar, starting from the year in which the transformer architecture \cite{transformer} was first published, \textit{i.e.} 2017, up until the date of publication of this article, \textit{i.e.} 2023, shown in Figure~\ref{fig:graphcompare} It becomes clear how quickly interest in this research topic has grown, mostly attributed to the successes of commercial large language models such as GPT-4, BERT and LLaMA. However, the amount of research which also includes \textit{"cybersecurity"} is lagging far behind, demonstrating not only a worrying trend of lack of oversight into such systems but also a gap of applying this new technology to the cybersecurity context.

% Remove resizebox if figure stays in one page.

It is important to note the distinction between \textit{"LLMs \textbf{in} cybersecurity"} and \textit{"cybersecurity \textbf{of} LLMs"}, with the former being the focus of this survey.

A survey of papers leveraging LLMs in the context of cybersecurity was carried by searching for the keywords \textit{"Large Language Models" AND "Cybersecurity"} and further selecting papers which used LLMs as a tool for cybersecurity in offensive or defensive contexts. An overview of the papers is shown in Table~\ref{tab:cybersecurity_publications}. The highlighted papers were selected for further analysis due to their increased applicability as autonomous vulnerability assessment agents in the cybersecurity context.

\renewcommand{\arraystretch}{1.3}
\begin{landscape}
\begin{longtable}{p{15cm} p{3.7cm} p{3.5cm}}
%\centering

\caption{Summary of Publications on Large Language Models in Cybersecurity} \label{tab:cybersecurity_publications} \\ % Your caption and label go here

\toprule
\centering\arraybackslash \textbf{Title}                                                                                                                                                                 & \centering\arraybackslash \textbf{Approach}                                        & \centering\arraybackslash \textbf{Model}                                                                                          \\
\midrule
\textbf{Using Large Language Models for Cybersecurity Capture-The-Flag Challenges and Certification Questions \cite{tannUsingLargeLanguage2023}}                               & \textbf{CTF Solving}                                     & \textbf{GPT-3.5; PaLM 2; Prometheus}                                     \\
An Empirical Study on Using Large Language Models to Analyze Software Supply Chain Security Failures \cite{singlaEmpiricalStudyUsing2023}                             & Software supply chain breach analysis           & ChatGPT-3.5-turbo; LaMDA                                                      \\
\textbf{Out of the Cage: How Stochastic Parrots Win in Cyber Security Environments \cite{rigakiOutCageHow2023}}                                                                & \textbf{Network pentesting}                              & \textbf{GPT-3.5-turbo; GPT-4}                                                                           \\
\textbf{Examining Zero-Shot Vulnerability Repair with Large Language Models \cite{pearceExaminingZeroShotVulnerability2023}}                                                   & \textbf{Source code vulnerability fixing}                & \textbf{code-cushman-001; code-davinci-001; code-davinci-002; j1-jumbo; j1-large; polycoder; gpt2-csrc} \\
VulDefend: A Novel Technique based on Pattern-exploiting Training for Detecting Software Vulnerabilities Using Language Models \cite{omarVulDefendNovelTechnique2023} & Source code vulnerability detection             & RoBERTa                                                                                        \\
Harnessing GPT-4 for Generation of Cybersecurity GRC Policies: A Focus on Ransomware Attack Mitigation \cite{mcintoshHarnessingGPT4Generation2023}                    & GRC policy writing                              & GPT-4                                                                                          \\
Detecting Phishing Sites Using ChatGPT \cite{koideDetectingPhishingSites2023}                                                                                         & Phishing website detection                      & GPT 3.5; GPT-4                                                                                 \\
ChatIDS: Explainable Cybersecurity Using Generative AI \cite{juttnerChatIDSExplainableCybersecurity2023a}                                                             & IDS alert explainability                        & GPT-3.5-turbo                                                                                  \\
Devising and Detecting Phishing: large language models vs. Smaller Human Models \cite{heidingDevisingDetectingPhishing2023}                                           & Phishing email detection                        & GPT-4; Claude; PaLM; LLaMA                                                                     \\
Large Language Models Can Be Used To Effectively Scale Spear Phishing Campaigns \cite{hazellLargeLanguageModels2023}                                                  & Spear phishing email creation                   & GPT-3.5; GPT-4                                                                                 \\
\textbf{Getting pwn'd by AI: Penetration Testing with Large Language Models \cite{happeGettingPwnAI2023}}                                                                      & \textbf{Penetration testing and vulnerability detection} & \textbf{GPT-3.5-turbo}                                                                                  \\
Revolutionizing Cyber Threat Detection with Large Language Models \cite{ferragRevolutionizingCyberThreat2023}                                                         & Threat detection and response                   & SecureBERT                                                                                     \\
SecureFalcon: The Next Cyber Reasoning System for Cyber Security \cite{ferragSecureFalconNextCyber2023}                                                               & Source code vulnerability detection             & FalconLLM                                                                                      \\
Prompting Is All You Need: Automated Android Bug Replay with Large Language Models \cite{fengPromptingAllYou2023}                                                     & Android bug replay                              & GPT-3.5                                                                                        \\
On the Uses of Large Language Models to Interpret Ambiguous Cyberattack Descriptions \cite{fayyaziUsesLargeLanguage2023}                                              & TTP interpretation                              & BERT; SecureBERT                                                                               \\
\textbf{PentestGPT: An LLM-empowered Automatic Penetration Testing Tool \cite{dengPentestGPTLLMempoweredAutomatic2023}}                                                        & \textbf{Penetration testing}                            & \textbf{GPT-3.5; GPT-4; LaMDA}                                                                           \\
From Text to MITRE Techniques: Exploring the Malicious Use of Large Language Models for Generating Cyber Attack Payloads \cite{charanTextMITRETechniques2023}         & Malicious payload generation                    & GPT-3.5; LaMDA                                                                                  \\
RatGPT: Turning online LLMs into Proxies for Malware Attacks \cite{beckerichRatGPTTurningOnline2023}                                                                  & Attack proxy                                    & GPT-3.5                                                                                        \\
Transformer-Based Language Models for Software Vulnerability Detection \cite{thapaTransformerBasedLanguageModels2022}                                                 & Source code vulnerability detection             & BERT; DistilBERT; RoBERTa; CodeBERT; GPT-2; MegatronBERT; MegatronGPT-2; GPT-J                 \\
VulRepair: a T5-based automated software vulnerability repair \cite{fuVulRepairT5basedAutomated2022}                                                                  & Source code vulnerability fixing                & CodeT5                                                                                         \\ \bottomrule
\end{longtable}
\end{landscape}

\subsection{PentestGPT: An LLM-empowered Automatic Penetration Testing Tool}

The study first builds on current base models (GPT-3.5, GPT-4 and Bard) through a iterative approach, i.e. passing information to and from the tested back and forth and prompting the LLM, in order to test their usability in a pentesting scenario. The authors find that the LLMs are able to provide a good "intuition" on how to proceed with the task, however, a major drawback is that context gets lost as prompting advances and the system loses its ability to correctly probe and decide on further tasks.

In order to solve these issues, the authors propose a framework with tooling and benchmarks, with its most significant constituent being the "PentestGPT", an LLM with reasoning, generation and parsing modules. Each module represents a role within a red team and was developed through Chain-of-Thought prompting that defined its role on the commercially available models GPT-3.5 and GPT-4. As shown in benchmarking, the "PentestGPT-GPT4" approach presented the best results.

\subsection{Getting pwn'd by AI: Penetration Testing with Large Language Models}

Two approaches were taken in this study: as a "sparring" partner for penetration testing, in which the LLM acts in a Q\&A fashion; and as an automated agent connected to a virtual machine (VM) in a scenario where the pentester already got initial access and is performing further privilege escalation.

The LLM performed well in the Q\&A approach and replied with realistic and feasible suggestions in order to perform the initial pentesting, although some filtering which is present in commercial models sometimes impacted on the results.

For the second approach, a script was made that kept the model in a loop with the VM so that it could issue commands and get the output from the terminal. The model was successfully able to gain root privilege multiple times within the vulnerable machine, but failed to perform multi-step exploitation. Besides the previous issue with filtering, this approach also had hallucinations in which the model tried to run scripts which did not exist in the environment, although the authors didn't find this issue occurred too often.

\subsection{Examining Zero-Shot Vulnerability Repair with Large Language Models}

The study compares several commercial foundation models already pre-trained on code tasks. Zero-shot refers to direct prompting of the model, in comparison with few-shot in which the prompt is engineered with a few examples. Because the models used were already trained mostly on code, they are better suited for code tasks with fewer prompting.

The authors found that although successful, most commercial models struggle with more complex tasks, and results were greatly dependent on the quality of the prompt. They also trained their own local model which, due to its reduced size, did not perform as well as the commercial models. However, in general, the models were able to fix vulnerable code, even if it took more than one attempt.

\subsection{Out of the Cage: How Stochastic Parrots Win in Cyber Security Environments}

The authors used commercial, pre-trained models in order to create single-prompt agents. The agents were only prompted with instructions and rules of a game that was developed as a reinforcement learning playground and state and memory information plus a small prompt of examples of valid actions and a query for the next step. They found that memory and temperature (which corresponds to a "randomness" toggle) were important factors so that the agent wouldn't get stuck. There were also hallucination issues, in which the agent would suggest impossible actions.

The latest commercial model, GPT-4, also performed significantly better than GPT-3.5-turbo, and had better reasoning skills so it would get stuck less often.

\subsection{Using Large Language Models for Cybersecurity Capture-The-Flag Challenges and Certification Questions}

The study focuses on two scenarios of using LLMs: Q\&A at the certification level; and CTF challenge solving.

In the first scenario, the LLM performed better in factual questions than conceptual questions. The authors tested on Cisco certifications ranging from the associate to the expert level.

For the second scenario, the authors prompted the models with CTF challenges. Three commercially available models, namely GPT-3.5, PaLM 2 and Prometheus, were tested. A back-and-forth conversation was taken, in which the authors prompted the models describing the CTF challenge, asking for further direction and replying with feedback from exploring the challenge.

All models were successful in aiding with the CTF challenges, although one major limitation was security filtering, which was bypassed through jailbreaks.

\section{Current issues and challenges}

Although current state-of-the-art LLMs excel at generating small to medium text snippets, the quality of the generated output tends to quickly degrade as the conversation evolves and when dealing with complex tasks where additional context may "leak" into the prompt. Notably, this issue is present even in models with large context windows \cite{shi2023large}. Similarly, the way in which context is given to a model and even where relevant information is located, \textit{e.g.} at the beginning of the query \textit{vs} at the end \textit{vs} repeated throughout, may change the quality of the output provided by the model \cite{liu2023lost}.

This \textbf{loss of context} becomes especially challenging when trying to provide agency for a system where the inputs and outputs can't always be controlled and improved by a human supervisor, leading to a quick degradation of the model on the performed task.

Another common issue with LLMs are \textbf{hallucinations}, which happens when the models make up content not grounded in reality or contradictory with previous content and context \cite{mündler2023selfcontradictory, dziri2022origin, Ji_2023, zhang2023sirens, maynez2020faithfulness}.

These are harder to tackle and represent one of the most significant issue of current models. Some hallucinations are harder to pinpoint as they can simply be considered bias from the dataset. This behaviour may also be desirable in some cases, such as for models in which creative generation is the desired output instead.

The current most common approaches for mitigating these issues are given bellow:

\subsection{Fine-tuning}

Because current models are prohibitively expensive to fully train when new data becomes available, a further post-training step may be carried \cite{zhang2023instruction}. This also allows adding private data into the models, such as internal company documents and personal information \cite{Behnia_2022}.

Several approaches are being researched in order to perform fine-tuning in a more efficient way \cite{huLORALOWRANKADAPTATION, dettmersQLoRAEfficientFinetuning2023, malladiFineTuningLanguageModels2023, chen2023longlora, Ding2023, lv2023parameter} and although it is orders of magnitude cheaper than training from scratch, its costs still remains high.

A more significant issue is catastrophic forgetting, in which the model loses some of it original capability after fine-tuning \cite{lin2023speciality, luo2023empirical, zhai2023investigating}.

\subsection{In-Context Learning (ICS)/Few-shot learning}

By providing the model with a few examples of how its expected behavior should look like, the model tends to perform better than if the examples were not displayed to it \cite{wei2023larger, brownLanguageModelsAre2020}.

This technique is striking due to its simplicity as well as its effectiveness and can somewhat approximate the fine-tuning of a model without the more computationally intensive training step, providing instead a guidance towards the desired behavior.

However, hallucinations are still common and the examples provided via the prompt end up consuming tokens that could instead be used with information. This decreases the effective prompting window.

\subsection{Retrieval Augmented Generation (RAG)}

By combining the LLMs with a vector database, which groups tokens by semantic proximity, it is possible to merge the results of the model generation with the contents of the database which are grounded in factual information \cite{NEURIPS2020_6b493230, 10.1145/3477495.3532682, mao2021generationaugmented, li2022survey, liu2021retrievalaugmented}.

This technique is especially relevant as it can also provide citations on where from the database the information comes from. It still has high costs, as the information must be processed and stored in the specialized database. There is also an issue in which the retrieved information may be semantically related but not always factually relevant.

\subsection{Chain-of-Verification (CoVe)}

This approach involves tasking the model with creating a first draft of the response; questioning its own output and fact-checking its responses independently, \textit{i.e.} without context of the draft in order to avoid bias; and, finally, generating a final, verified response \cite{dhuliawala2023chainofverification}. Although the quality of the responses increases and the hallucination is greatly reduced, it does not completely solve the hallucinations and increases the computational load significantly as each query needs to be repeated with the verification prompt as well.

\section{Proposed solution}

\textbf{Mixture-of-Experts} (MoE) is a technique already used in previous deep learning systems \cite{6215056} which is making its way into current state-of-the-art LLM models \cite{zhou2022mixtureofexperts, puigcerver2023sparse} and allows for high quality inference without the high costs associated with other techniques.

The step-by-step overview of this approach is as follows:
\begin{labeling}{\textbf{Subtask identification}}
    \item[\textbf{Subtask identification}] The main task is subdivided into subtasks which may be addressed independently;
    \item[\textbf{Expert modelling}] Each subtask is assigned to an expert model;
    \item[\textbf{Gating modelling}] A model is trained to recognize the initial task and to which expert model it should be routed to;
    \item[\textbf{Combination}] the output is produced by combining the results of the experts with the gating model.
\end{labeling}

The proposed solution is a Mixture-of-Experts-based approach in which each expert is a task-specific foundation model. Foundation models are usually trained for a desired task, such as instruction following, conversation and code generation. The combination of specialized models may lead to reduced hallucinations, better context management and better specificity on the results.

This technique is particularly well suited for vulnerability assessment tasks given it requires a wide ranging skill set.

The approach is similar to \textbf{PentestGPT} \cite{dengPentestGPTLLMempoweredAutomatic2023}, in which three different modules are presented: Reasoning, Generation and Parsing. Each module has its own set of tasks, however, specialization is acquired merely through keeping each module with a narrow task-space at inference level, \textit{i.e.} the model themselves are not specially trained. Although not as in-depth, this approach already manages to be effective.

We propose greater specialization and model-localization by leveraging different foundation LLMs for each task as expert models, with one acting as the gating model which chooses the expert for the task at hand.

Instruction-tuned models \cite{ouyang2022training} are ideal for the reasoning aspects of the tasks as they are already aligned with sentences structured as instructions.

Models specialized in code \cite{rozière2023code} may then receive instructions to interpret one section of code, find vulnerabilities or develop exploits which can then loop back to the other models as a response or tool to be used.

The gating model should be trained on the widest range of subjects so it can understand the task given and to which expert model it should be delegated, \textit{e.g.} it needs to understand what code is, but it doesn't need to excel at coding. These can be seen as the "generic", commercial foundation models \cite{touvronLlamaOpenFoundation2023, llama, openaiGPT4TechnicalReport2023}.

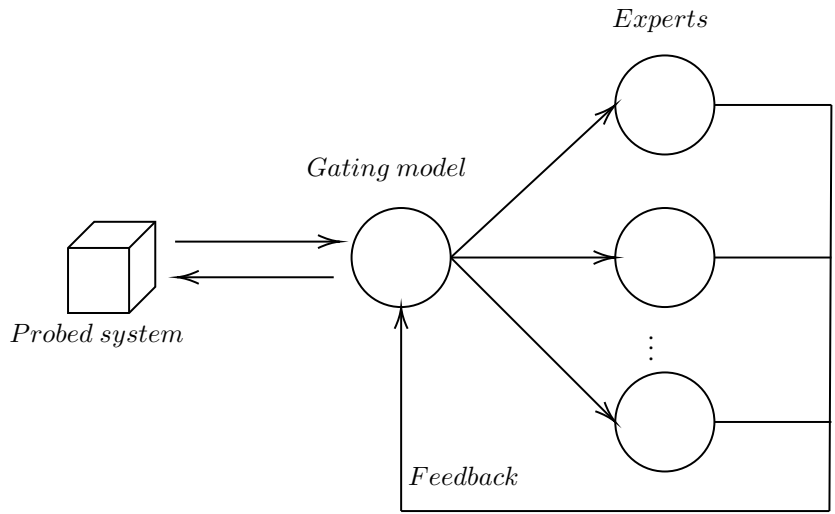
\begin{figure}[h!]
    \centering
    \tikzset{every picture/.style={line width=0.75pt}} %set default line width to 0.75pt        

\begin{tikzpicture}[x=0.75pt,y=0.75pt,yscale=-1,xscale=1]
%uncomment if require: \path (0,455); %set diagram left start at 0, and has height of 455

%Shape: Cube [id:dp7435297878640204] 
\draw   (136,277.17) -- (149.2,263.97) -- (180,263.97) -- (180,296.8) -- (166.8,310) -- (136,310) -- cycle ; \draw   (180,263.97) -- (166.8,277.17) -- (136,277.17) ; \draw   (166.8,277.17) -- (166.8,310) ;
%Shape: Circle [id:dp5273998071086865] 
\draw   (279,282) .. controls (279,268.19) and (290.19,257) .. (304,257) .. controls (317.81,257) and (329,268.19) .. (329,282) .. controls (329,295.81) and (317.81,307) .. (304,307) .. controls (290.19,307) and (279,295.81) .. (279,282) -- cycle ;
%Straight Lines [id:da494277596700599] 
\draw    (190,273.97) -- (271,273.97) ;
\draw [shift={(273,273.97)}, rotate = 180] [color={rgb, 255:red, 0; green, 0; blue, 0 }  ][line width=0.75]    (10.93,-3.29) .. controls (6.95,-1.4) and (3.31,-0.3) .. (0,0) .. controls (3.31,0.3) and (6.95,1.4) .. (10.93,3.29)   ;
%Straight Lines [id:da2118089293437957] 
\draw    (270,291.97) -- (193,291.97) ;
\draw [shift={(191,291.97)}, rotate = 360] [color={rgb, 255:red, 0; green, 0; blue, 0 }  ][line width=0.75]    (10.93,-3.29) .. controls (6.95,-1.4) and (3.31,-0.3) .. (0,0) .. controls (3.31,0.3) and (6.95,1.4) .. (10.93,3.29)   ;
%Shape: Circle [id:dp24156931091908995] 
\draw   (412,282) .. controls (412,268.19) and (423.19,257) .. (437,257) .. controls (450.81,257) and (462,268.19) .. (462,282) .. controls (462,295.81) and (450.81,307) .. (437,307) .. controls (423.19,307) and (412,295.81) .. (412,282) -- cycle ;
%Straight Lines [id:da11808635997973482] 
\draw    (329,282) -- (410,282) ;
\draw [shift={(412,282)}, rotate = 180] [color={rgb, 255:red, 0; green, 0; blue, 0 }  ][line width=0.75]    (10.93,-3.29) .. controls (6.95,-1.4) and (3.31,-0.3) .. (0,0) .. controls (3.31,0.3) and (6.95,1.4) .. (10.93,3.29)   ;
%Shape: Circle [id:dp8873527891082147] 
\draw   (412,205) .. controls (412,191.19) and (423.19,180) .. (437,180) .. controls (450.81,180) and (462,191.19) .. (462,205) .. controls (462,218.81) and (450.81,230) .. (437,230) .. controls (423.19,230) and (412,218.81) .. (412,205) -- cycle ;
%Shape: Circle [id:dp38625321720263495] 
\draw   (412,365) .. controls (412,351.19) and (423.19,340) .. (437,340) .. controls (450.81,340) and (462,351.19) .. (462,365) .. controls (462,378.81) and (450.81,390) .. (437,390) .. controls (423.19,390) and (412,378.81) .. (412,365) -- cycle ;
%Straight Lines [id:da03383199212018129] 
\draw    (329,282) -- (410.53,206.36) ;
\draw [shift={(412,205)}, rotate = 137.15] [color={rgb, 255:red, 0; green, 0; blue, 0 }  ][line width=0.75]    (10.93,-3.29) .. controls (6.95,-1.4) and (3.31,-0.3) .. (0,0) .. controls (3.31,0.3) and (6.95,1.4) .. (10.93,3.29)   ;
%Straight Lines [id:da8483876527833794] 
\draw    (329,282) -- (410.59,363.59) ;
\draw [shift={(412,365)}, rotate = 225] [color={rgb, 255:red, 0; green, 0; blue, 0 }  ][line width=0.75]    (10.93,-3.29) .. controls (6.95,-1.4) and (3.31,-0.3) .. (0,0) .. controls (3.31,0.3) and (6.95,1.4) .. (10.93,3.29)   ;
%Straight Lines [id:da3947975146592171] 
\draw    (304,409.97) -- (304,309) ;
\draw [shift={(304,307)}, rotate = 90] [color={rgb, 255:red, 0; green, 0; blue, 0 }  ][line width=0.75]    (10.93,-3.29) .. controls (6.95,-1.4) and (3.31,-0.3) .. (0,0) .. controls (3.31,0.3) and (6.95,1.4) .. (10.93,3.29)   ;
%Straight Lines [id:da19704793225525163] 
\draw    (304,409.97) -- (520,409.97) ;
%Straight Lines [id:da5139530223507331] 
\draw    (462,205) -- (521,205) ;
%Straight Lines [id:da1600797713407882] 
\draw    (462,282) -- (521,282) ;
%Straight Lines [id:da28275576749953024] 
\draw    (462,365) -- (521,365) ;
%Straight Lines [id:da15243464386833372] 
\draw    (520,409.97) -- (521,205) ;

% Text Node
\draw (427,312) node [anchor=north west][inner sep=0.75pt]    {$\vdots $};
% Text Node
\draw (105,314) node [anchor=north west][inner sep=0.75pt]    {$Probed\ system$};
% Text Node
\draw (255,230) node [anchor=north west][inner sep=0.75pt]    {$Gating\ model$};
% Text Node
\draw (409,155) node [anchor=north west][inner sep=0.75pt]    {$Experts$};
% Text Node
\draw (306,386) node [anchor=north west][inner sep=0.75pt]    {$Feedback$};

\end{tikzpicture}
    \caption{Diagram of the proposed system.}
    \label{fig:enter-label}
\end{figure}

\section{Final remarks}

In conclusion, the Mixture-of-Experts framework presented in this paper represents a significant stride towards enhancing cybersecurity practices, leveraging the remarkable capabilities of Large Language Models (LLMs) for pen testing and vulnerability assessment. By orchestrating the collaboration of specialized LLMs, each adept in distinct cybersecurity domains, we have introduced a scalable and adaptable solution that holds promise in addressing the multifaceted challenges posed by modern cyber threats. The system's ability to allocate specific tasks to LLMs specializing in code analysis, reasoning, and anomaly detection ensures a comprehensive evaluation of security postures.

The implications of this research extend beyond theoretical frameworks, offering tangible prospects for fortifying digital infrastructures. The adaptability of the Mixture-of-Experts architecture positions it as a dynamic tool that can evolve alongside the ever-changing cybersecurity landscape. Nevertheless, it is imperative to acknowledge the current limitations, such as the dependence on the quality of LLMs and the necessity for comprehensive training datasets. Practical validation in diverse environments is essential to substantiate the efficacy of the proposed framework in real-world cybersecurity scenarios.

As the cybersecurity community grapples with increasingly sophisticated threats, the collaborative intelligence harnessed by the Mixture-of-Experts framework provides a foundation for future innovations. Through continuous refinement and validation, this approach has the potential to not only augment existing security measures but also lay the groundwork for intelligent, adaptive defense mechanisms against emerging digital risks.

\section{Future directions}

While this research has laid the groundwork for integrating LLMs into cybersecurity practices, numerous avenues for future exploration emerge. To further enhance the Mixture-of-Experts framework, a concerted effort is needed to refine the expertise of specialized LLMs. This involves continuous fine-tuning, leveraging domain-specific datasets, and exploring techniques to mitigate biases that may arise during the training process.

Expanding the scope of the framework to encompass additional cybersecurity domains is paramount. Future research should explore the integration of LLMs specialized in threat intelligence, incident response, and even regulatory compliance. This expansion will contribute to a more holistic and multifaceted approach to cybersecurity, addressing not only vulnerabilities and threats but also the broader landscape of risk management and compliance.

Empirical validation remains a crucial aspect of future research endeavors. Conducting extensive experiments in diverse and realistic cybersecurity environments will provide valuable insights into the practical effectiveness of the Mixture-of-Experts framework. This includes evaluating the framework's performance in scenarios involving sophisticated adversaries, varied network architectures, and different industry sectors.

Furthermore, as ethical considerations surrounding AI and cybersecurity continue to gain prominence, future research should delve into the development of responsible AI frameworks within the proposed system. Ensuring transparency, accountability, and fairness in decision-making processes will be essential for gaining trust in the deployment of such intelligent systems within critical security infrastructures.

In summary, the future trajectory of this research should focus on refining, expanding, and validating the Mixture-of-Experts framework, paving the way for a new era of intelligent and collaborative cybersecurity defenses. As we navigate the complexities of the digital landscape, the ongoing development of advanced frameworks will be instrumental in safeguarding our interconnected world against evolving cyber threats.

\newpage

%Bibliography
\bibliographystyle{unsrt}  
\bibliography{references}

\end{document}